\documentclass[11pt,a4paper]{article}
\usepackage{jheppub_kim}
\usepackage{subfigure,amsmath,amssymb ,amsfonts, latexsym}
\usepackage[notcite,color,final]{showkeys}

\title{Constraints on a $\phi$CDM model from strong gravitational lensing and updated Hubble parameter measurements}
\author{Yun Chen$^{a,b,d}$, Chao-Qiang Geng$^{a,b}$\footnote{Corresponding Author: geng@phys.nthu.edu.tw}, Shuo Cao$^c$, Yu-Mei Huang$^c$,  Zong-Hong Zhu$^c$}

\def\aa{Astron. \& Astrophys.}
\def\aar{Astron. \& Astrophys.~Rev.}
\def\aj{Astron. J.~}
\def\apj{Astrophys. J.~}

\def\araa{Ann. Rev. Astron. Astrophys.~}

\def\jcap{J.~Cosmol.~Astropart.~Phys.~}

\def\plb{Phys. Lett. B~}
\def\prl{Phys. Rev. Lett.~}
\def\prd{Phys. Rev. D~}

\def\mnras{Mon. Not. R. Astron. Soc.~}

\affiliation[a]{ Department of Physics, National Tsing Hua University,
Hsinchu, Taiwan 300}

\affiliation[b]{National Center for Theoretical Sciences, Hsinchu,
Taiwan 300}

\affiliation[c]{Department of Astronomy, Beijing Normal
University, Beijing 100875, China}

\affiliation[d]{Key Laboratory for Computational Astrophysics, National Astronomical Observatories, Chinese Academy of Sciences, Beijing, 100012, China}

\abstract
{
We constrain the scalar field dark energy model with an inverse power-law potential, i.e.,
$V(\phi)\propto {\phi}^{-\alpha}$ ($\alpha>0$), from a set of  recent cosmological observations by compiling
an updated sample of Hubble parameter measurements including 30 independent data points.
Our results show that the constraining power of the updated sample of $H(z)$ data with the HST prior
on $H_0$ is stronger than those of the SCP Union2 and Union2.1 compilations. A recent sample of strong gravitational lensing systems is also  adopted to confine the model even though  the results  are not
significant.  A joint analysis of the strong gravitational lensing data with the more restrictive updated
Hubble parameter measurements and the Type Ia supernovae data from SCP Union2 indicates that
the recent observations still can not distinguish whether dark energy is a time-independent cosmological
constant or a time-varying  dynamical component.
}


\begin{document}

\maketitle

\section{Introduction}
\label{Intro}
The accelerating expansion of the universe, which was first discovered in 1998 by two groups independently
via type Ia supernovae (SNeIa)~\cite{Riess1998, Permutter1999}, is a big challenge to the standard paradigms
of the fundamental physics and cosmology.  A number of approaches have been adopted to explain this remarkable
discovery~\cite{Copeland2006, Ratra&Vogeley2008, LiM2011}.
In one of the scenarios, an exotic component dubbed dark energy, which is expected to possess a negative pressure,
is proposed to understand the accelerating expansion. The simplest candidate of dark energy
is the $\Lambda$CDM model, which is the current standard model of cosmology and in good agreement with most
observations~\cite{Frieman2008, Jassal2010}. However, the $\Lambda$CDM model is embarrassed by two
problems~\cite{Weinberg1989}, namely, the fine-tuning~\cite{Carroll2001} and coincidence~\cite{Zlatev1999, ChenY2010}
problems. Nevertheless, there is a different view which shows that the so called ``coincidence problem'' is just an artifact of
the anthropic selection~\cite{Sivanandam2013}. Up to now, there are mainly three ways to explain or alleviate
the cosmological constant problems, i.e., the anthropic principle~\cite{Weinberg2000},
the dynamical dark energy~\cite{Copeland2006}, and the interacting
dark energy model~\cite{Amendola2000, Caldera-Cabral2009}. Quintessence is one of the hypothetical forms of
the dynamical dark energy, which is described by a scalar field $\phi$ minimally coupled to gravity.
With some particular form of the potential $V(\phi)$, it can lead to the late time inflation-like accelerating
cosmological expansion.  An incomplete list of $V(\phi)$ is shown in Table 3.1 of Ref.~\cite{Samushia2009}.
Quintessence models with some particular potentials can lighten the cosmological
coincidence problem~\cite{Zlatev1999}. The inverse power-law potential
$V(\phi)\propto \phi^{-\alpha}$ ($\alpha>0$)~\cite{Peebles&Ratra1988} is one of this kind of the potentials.
In this paper, we focus on the quintessence model with the inverse power-law potential.

Currently, type Ia supernovae (SNe Ia),  baryon acoustic oscillations (BAO) and
cosmic microwave background (CMB) are deemed as three types of standard cosmological probes
to give strict constraints on the cosmological parameters~\cite{Planck2013}.
However, it is necessary and important to have other complementary cosmological probes,
including the Hubble parameter measurements, the strong gravitational lensing, the galaxy cluster gas mass fraction,
the gamma-ray burst, the angular diameter distance data of galaxy clusters, and the radio galaxy angular size data.
In this paper, we will mainly investigate the constraints on the inverse power-law $\phi$CDM model from an updated
sample of the Hubble parameter measurements as well as the recent strong gravitational lensing (SGL) data.

The paper is organized as follows. In Sec.~\ref{ModelReview}, we review quintessence models of dark energy,
mainly with an inverse power-law potential, i.e., $V(\phi)\propto \phi^{-\alpha}$ ($\alpha>0$).
In Sec.~\ref{ObsConstr}, we examine constraints on the model from different observational samples,
particularly updated $H(z)$ and recent SGL data sets.
 Finally, we discuss our results and present our main conclusions in Sec.~\ref{summary}.

\section{$\phi$CDM model with the inverse power-law potential}
\label{ModelReview}

Scalar fields, which naturally arise in particle physics can act as candidates for dark energy to explain
the late-time cosmic acceleration. So far, there have been many scalar field dark energy models,
such as quintessence~\cite{Carroll1998, Tsujikawa2013}, phantom~\cite{Caldwell2002},
k-essence~\cite{Chiba2000, Chiba2009}, and tachyon~\cite{Sen2002}.
Among them, quintessence, which is canonical, is probably the most popular one.

We consider the self-interacting quintessence minimally coupled to gravity.
Since quintessence is described by the scalar field $\phi$, the corresponding dark energy models
can also be called as $\phi$CDM models. In what follows, we will use the terms ``quintessence''
and ``$\phi$CDM'' essentially interchangeably. The action of the models is given by
\begin{equation}
\label{eq:totAction}
S=\int~\sqrt{-g}\left(-\frac{m^2_p}{16 \pi}R + \mathcal{L}_{\phi}+\mathcal{L}\right)d^4 x,
\end{equation}
where $g$ is the determinant of the metric $g_{\mu\nu}$, $R$ is the Ricci scalar,
$m_p=1/\sqrt{G}$ is the Planck mass with $G$ being the Newtonian constant of gravitation,
$\mathcal{L}$ is the Lagrangian density for matter and radiation, and $\mathcal{L}_{\phi}$
is the Lagrangian density for the field $\phi$, given by
\begin{equation}
\mathcal{L}_{\phi}=\frac{m^{2}_{p}}{16 \pi}\left[\frac{1}{2}g^{\mu \nu}\partial_{\mu}\phi\partial_{\nu}\phi-V(\phi)\right].
\label{eq:Lphi}
\end{equation}
In this work, we take a flat Friedmann-Lema\^{i}tre-Robertson-Walker (FLRW) universe.
In the framework, by the variation of the action in
Eq.~(\ref{eq:totAction}) with respect to $\phi$, one can derive the equation of motion for $\phi$, given by
\begin{eqnarray}
\ddot\phi+3\left(\frac{\dot a}{a}\right)\dot \phi+\frac{d V}{d\phi}=0,
\label{eq:KGeq}
\end{eqnarray}
which is just the Klein-Gordon equation.
In theories beyond the standard model of particle physics, such as
string theory, minimal supersymmetric theory, etc., scalar fields arise naturally.
However, since the underlying physics is not known at present, it is impossible to set a specific functional
form for $V(\phi)$ yet.
Additionally, for quintessence models, there are many kinds of $V(\phi)$ which can satisfy
the requirement of the late-time accelerating expansion of the universe~\cite{Samushia2009}.
In 1988, Peebles and Ratra~\cite{Peebles&Ratra1988} proposed a scalar field that is slowly rolling down
with a potential  $V(\phi)=\frac{1}{2}\kappa m^{2}_{p}\phi^{-\alpha}$ at a large $\phi$,
where $\kappa$ and $\alpha$ are non-negative parameters. This kind of the inverse power-law potential
can not only lead to the late-time acceleration of the universe but also partially solve the two cosmological
constant problems mentioned in Sec.~\ref{Intro}. When $\alpha = 0$, this $\phi$CDM model is reduced to
the $\Lambda$CDM case. In this paper, we concentrate on this type of quintessence models.

The Friedmann equation of the $\phi$CDM model can be written as
\begin{equation}
\label{eq:phiCDMFriedmann} H^2(z) =
\frac{8\pi}{3m_p^2}(\rho_m + \rho_{\phi}),
\end{equation}
where $H(z) \equiv \dot{a}/a$ is the Hubble parameter.
The energy densities are given by
\begin{equation}
\label{eq:rhom} \rho_m = \frac{m_p^2}{6\pi}a^{-3},
\end{equation}
and
\begin{equation}
\label{eq:rhophi} \rho_{\phi} = \frac{m_p^2}{32\pi} (\dot{\phi}^2 +
\kappa m_p^2 \phi^{-\alpha}).
\end{equation}
Based on Eqs.~(\ref{eq:KGeq}) and (\ref{eq:phiCDMFriedmann}), along with the initial conditions described in
Ref.~\cite{Peebles&Ratra1988}, one can numerically compute the Hubble parameter $H(z)$.
The parameter $\kappa$ depends on $\alpha$ through the following equality~\cite{{FarooqPhDthesis}},
\begin{eqnarray}
\kappa m^2_p=\frac{8(\alpha+4)}{3(\alpha+2)}\left[\frac{2}{3}\alpha(\alpha+2)\right]^{\alpha/2}.
\label{eq:KappaAlpha}
\end{eqnarray}
According to  the definition of the dimensionless
density parameter, one has
\begin{equation}
\label{eq:Omegam} \Omega_m(z) = \frac{8\pi \rho_m}{3m_p^2H^2} =
\frac{\rho_m}{\rho_m + \rho_{\phi}}.
\end{equation}
In this case, the
model-parameter set is $\textbf{p} = (\Omega_{m0}, \alpha)$, where $\Omega_{m0}=\Omega_m(z=0)$.

\section{Constraints from different observational data sets}
\label{ObsConstr}
\subsection{The updated Hubble parameter measurements}

 An approach was proposed  by Jimenez and Loeb (2002)
 in Ref.~\cite{JimenezLoeb2002}, which proposes that the passively evolving galaxies
 can be used as standard cosmic chronometers whose differential age evolution as a function of the redshift
 can directly probe $H(z)$.
 The expansion rate, i.e. the Hubble parameter, is defined as:
 \begin{equation}
\label{eq:Hz_DA} H(z)=\frac{\dot{a}}{a}=-\frac{1}{1+z}\frac{dz}{dt}\,,
\end{equation}
where the redshift $z$ of the chronometers is known with a high accuracy based on the spectra of galaxies and
the differential measurement of time ($dt$) at a given redshift interval ($dz$) automatically provides a direct measurement on $H(z)$.
Hereafter, this \emph{differential age} approach will be quoted as ``DA'' for short.

Jimenez et al.~\cite{Jimenez_etal2003} obtained one $H(z)$ data point at $z=0.09$
as the first to use the ``DA'' method,
while Simon et al.~\cite{Simon2005} got 8 more $H(z)$ data up to $z = 1.75$.
Then, Stern et al.~\cite{Stern2010} improved over the $H(z)$ measurements in Refs.~\cite{Jimenez_etal2003, Simon2005} by including
two new determinations at $z = 0.48$ and $0.88$, and listed the total 11 measurements.
After that, Moresco et al.~\cite{Moresco2012} provided 8 new measurements of $H(z)$
by using both BC03~\cite{BC03} and MaStro~\cite{MaStro2011} stellar population synthesis models.
Recently, in Ref.~\cite{Zhang2012} four more $H(z)$ data were obtained with the ``DA'' approach.
The total 23 ``DA'' measurements of $H(z)$ from Refs.~\cite{Stern2010, Moresco2012, Zhang2012} are listed in
Table~\ref{tab:Hz}.
The $H(z)$ data within the same reference are statistically independent.
However, some of them from different references may have correlations
since the redshift intervals of the data to derive them have overlaps.
As there is no  canonical technique to estimate this kind of correlations,
we choose to ignore the effect and treat the 23 ``DA'' data as independent measurements.

The process of the ``DA'' method basically contains three steps.
The first step is to select an appropriate sample of passively evolving early-type galaxies (ETGs hereafter)
with high-quality optical spectra.
Note that the compilation of the data based on the ``DA'' method is in itself heterogeneous
 since each reference uses a different sample selection for its study.
 For example, Ref.~\cite{Stern2010} complemented the Keck spectra of red-envelope galaxies
 in 24 rich galaxy clusters with archival spectra from SDSS luminous red galaxies (LRGs) in DR6,
 as well as from SPICES and VVDS galaxies, while Ref.~\cite{Zhang2012} utilized the selected LRGs from SDSS DR7
 based on `Carson \& Nichol sample'~\cite{Carson2010}.

 The second step is to derive the age information from the spectra of galaxies.
 The stellar-population synthesis models  fit the observed spectra
 and then determine the age of the oldest stars in the selected galaxies (marginalizing over the metallicity).
 In order to investigate the systematic error arising in the age determination
 in the stellar population models, Ref.~\cite{Stern2010} took the mock spectra
 with the Charlot \& Bruzual (2008) (CB08) stellar-population models and
 then recovered the age (marginalizing over the metallicity) with the BC03 models,
 which rely on  the stellar interior
and atmosphere models. The results show that the good age recovery
 can be obtained with a random dispersion $\sim10\%$ for two different stellar-population models, CB08 and BC03,
 especially at higher metallicities which correspond to the typical metallicity of the old and massive galaxies.
 Note that those with both BC03
 and MaStro stellar population synthesis models in Ref.~\cite{Moresco2012}
 are also in good accordance.

 The third step is to compute differential ages at different redshifts and then  compute
 the $H(z)$ measurements with Eq.~(\ref{eq:Hz_DA}). The sample should be sub-divided into several redshift intervals,
 and the number of the intervals is the number of $H(z)$ measurements from the sample.
 The range of the redshift intervals should be not only small to reduce the statistical error, they have to be large enough to ensure that the age evolution in the redshift interval is larger than the error in the age determination.
 Subsequently, one can further fit each sub-sample by dividing it into suitably large redshift bins
 and acquire the $H(z)$ measurement in the given redshift interval at an effective redshift.
 While the redshift bins should be small to avoid incorporating
galaxies that have already evolved in ages, they also have to be large enough for the sparse sample to have more than one galaxy in most of the bins.
For more details on the selections of the redshift interval and binning, see Ref.~\cite{Simon2005}.

The systematic error of the ``DA'' method is related to the metallicity range spanned by the data, and the star formation history (SFH), as well as the adopted stellar population synthesis models, etc. From  Table 4 of Ref.~\cite{Moresco2012}, one can find out that the systematic error is about 1-10\%, while the total error is about 5-14\% in the redshift range $0.15 < z < 1.1$. More details on the systematic error are available from Refs.~\cite{Stern2010, Moresco2012}.

Another independent way of measuring $H(z)$ is from the clustering of galaxies or quasars.
Hereafter, this approach will be quoted as ``Clustering'' for short. It was first put forward
by Gazta{\~n}aga et al.~\cite{Gaztanaga2009} to get a direct measurement of $H(z)$ by
using the BAO peak position as a standard ruler in the radial direction.
Most detections of the BAO have used spherically averaged clustering statistics,
yielding a measurement of $D_V=D_A^{2/3}(cz/H(z))^{1/3}$ which is a combination of
the Hubble parameter $H(z)$ and  angular diameter distance $D_A(z)$. However,
by separating the clustering of the LRG sample in the SDSS DR6 and DR7 into the line-of-sight (LOS)
and transverse information, Ref.~\cite{Gaztanaga2009} directly inferred $H(z)$ from the LOS data.
Two independent measurements of $H(z)$ are found: $H(z=0.24)=79.69\pm2.32(\pm1.29)$ km/s/Mpc for $z=0.15-0.30$ and
$H(z = 0.43) = 86.45\pm3.27(\pm1.69)$ km/s/Mpc for $z = 0.40 - 0.47$, where the first and second errors are
the statistical and systematical errors, $\sigma_{stat}$ and  $\sigma_{sys}$, respectively.
 We also report the total error, estimated by summing in quadrature the statistical and systematic errors
 shown in Table~\ref{tab:Hz}. The systematic error mainly arises from the fitting methodology
 of determining the sound horizon from the galaxy correlation function, the magnification bias,
and the choices of the radial selection function and the angular mask as well as the model
used in the Monte Carlo simulations to estimate the statistical error.

Blake et al. in Ref.~\cite{Blake2012} took a more indirect route with the WiggleZ Dark Energy survey.
They first used the angle-averaged galaxy correlation function to determine the acoustic parameter
$A(z) \propto [D_A^2(z)/H(z)]^{1/3}$, and then fitted the 2D power spectrum to extract the Alcock-Paczynski
distortion parameter $F(z) \propto D_A(z) H(z)$. At last, by using the joint measurements of $A(z)$ and $F(z)$,
they broke the degeneracy between $D_A(z)$ and $H(z)$ and got the measurements of $D_A(z)$ and $H(z)$
at three effective redshifts. The corresponding measurements of $H(z)$ are $H(z=0.44)=82.6 \pm 7.8$ km/s/Mpc
for $z=0.2-0.6$, $H(z = 0.6) = 87.9 \pm 6.1$ km/s/Mpc for $z = 0.4 - 0.8$, and  $H(z = 0.73) = 97.3 \pm 7.0$ km/s/Mpc
for $z = 0.6 - 1.0$. Since the covariance matrix of the three $H(z)$ measurements is not available from Ref.~\cite{Blake2012},
we just employ the two independent measurements, i.e., $H(z=0.44)$ and  $H(z = 0.73)$.

Chuang and Wang in Ref.~\cite{Chuang2013} adopted a phenomenological model for the effective multipoles
of the correlation function and decomposed the BAO peak into radio and angular components to extract directly
$H(z)$ and $D_A(z)$ from the SDSS DR7 LRG sample, resulting in  $H(z = 0.35) = 79 \pm 12$ km/s/Mpc and
$H(z = 0.35) = 82.7 \pm 8.4$ km/s/Mpc with $40<s<120 h^{-1}$Mpc and $25<s<120 h^{-1}$Mpc, respectively.
As the constraints become tighter by including the smaller scales,
in Ref.~\cite{Chuang2013} they used those of $25<s<120 h^{-1}$Mpc as the fiducial results. In this work, we also take
$H(z = 0.35) = 82.7 \pm 8.4$ km/s/Mpc  with $25<s<120 h^{-1}$Mpc.

Anderson et al. in Ref~\cite{Anderson2013} measured $H(z)$ and $D_A(z)$ at the effective redshift $z=0.57$
from the SDSS III Baryon Oscillation Spectroscopic Survey (SDSS-III/BOSS) DR9  by applying the density-field reconstruction
to an anisotropic analysis of the BAO peak. The measurement of $H(z)$ is $H(z=0.57)=92.9 \pm 7.8$ km/s/Mpc,
where the systematic error  is below 1\%, which is negligible compared to the statistical one.

Busca et al. in Ref.~\cite{Busca2013} presented the first observation of the BAO peak by using the Ly$\alpha$ forest of BOSS quasars, which also gives the first BAO detection deep in the matter dominated epoch at $z=2.3$. Combined with the CMB constraints, they
found $H(z=2.3)=224 \pm 8$ km/s/Mpc.

The total 7 ``Clustering'' measurements of $H(z)$ from Refs.~\cite{Gaztanaga2009, Blake2012, Chuang2013,Anderson2013,Busca2013}
are listed in Table~\ref{tab:Hz}. In general, there are redshift interval overlaps for the 7 $H(z)$ data derived from clustering.
As these $H(z)$ measurements are generally derived from different surveys (SDSS, Wigglez) and  objects/observables
(LRGs, main galaxies, Ly$\alpha$ forest of quasars), these  data are assumed to be independent.
\begin{table*}
\caption{Data of the Hubble parameter $H(z)$ versus the redshift $z$, where $H(z)$ and $\sigma_{H}$ are in km s$^{-1}$
Mpc$^{-1}$.
}\label{tab:Hz}
\begin{center}{\scriptsize
\begin{tabular}{lrrll}
\hline
$z$ & $H(z)$ & $\sigma_{H}$ & Reference & Method \\
\hline
0.07 & 69.0 & 19.6 & \cite{Zhang2012} & DA\\
0.1 & 69.0 & 12.0 & \cite{Stern2010} & DA\\
0.12 & 68.6 & 26.2 & \cite{Zhang2012} & DA\\
0.17 & 83.0 & 8.0 & \cite{Stern2010} & DA \\
0.179 & 75.0 & 4.0 & \cite{Moresco2012} & DA\\
0.199& 75.0 & 5.0 & \cite{Moresco2012} & DA\\
0.2& 72.9 & 29.6 & \cite{Zhang2012} & DA\\
0.27 & 77.0 & 14.0 & \cite{Stern2010} & DA\\
0.28 & 88.8 & 36.6 & \cite{Zhang2012} & DA\\
0.352 & 83.0 & 14.0 & \cite{Moresco2012} & DA\\
0.4 & 95.0 & 17.0 & \cite{Stern2010} & DA\\
0.48 & 97.0 & 62.0 & \cite{Stern2010} & DA\\
0.593& 104.0 & 13.0 & \cite{Moresco2012} & DA\\
0.68 & 92.0 & 8.0 & \cite{Moresco2012} & DA\\
0.781 & 105.0 & 12.0 & \cite{Moresco2012} & DA\\
0.875 & 125.0 & 17.0 & \cite{Moresco2012} & DA\\
0.88 & 90.0 & 40.0 & \cite{Stern2010} & DA\\
0.9 & 117.0 & 23.0 & \cite{Stern2010} & DA\\
1.037 & 154.0 & 20.0 & \cite{Moresco2012} & DA\\
1.3 & 168.0 & 17.0 & \cite{Stern2010} & DA\\
1.43 & 177.0 & 18.0 & \cite{Stern2010} & DA\\
1.53 & 140.0 & 14.0 & \cite{Stern2010} & DA\\
1.75 & 202.0 & 40.0 & \cite{Stern2010} & DA\\
0.24 & 79.7 & 2.7 & \cite{Gaztanaga2009} & Clustering \\
0.35 & 82.7 & 8.4 & \cite{Chuang2013} & Clustering\\
0.43 & 86.5 & 3.7 & \cite{Gaztanaga2009} & Clustering\\
0.44 & 82.6 & 7.8 & \cite{Blake2012} & Clustering\\
0.57 & 92.9 & 7.8 & \cite{Anderson2013} & Clustering\\
0.73 & 97.3 & 7.0 & \cite{Blake2012} & Clustering\\
2.3  & 224.0 & 8.0 & \cite{Busca2013} & Clustering\\

\hline
\end{tabular}}\\
\end{center}
\end{table*}

 We constrain cosmological
parameters by minimizing $\chi_{H}^2$,
\begin{equation}
\label{eq:chi2Hz} \chi_{H}^2 (H_0, \textbf{p}) =
\sum_{i=1}^{N}\frac{[H^{\rm th} (z_i; H_0, \textbf{p})-H^{\rm
obs}(z_i)]^2}{\sigma^2_{{\rm H},i}},
\end{equation}
where $N$ is the number of data points, $z_i$ is the redshift at which $H(z_i)$ has been measured, $H^{\rm th}$
is the predicted value of $H(z)$ in the cosmological model and $H^{\rm obs}$ is the measured value. From
$\chi_{H}^2 (H_0, \textbf{p})$, we can compute the likelihood function
$\mathcal{L}_H(H_0, \textbf{p})$. We then treat $H_0$ as a nuisance parameter
and marginalize over it by using a gaussian prior with the mean $\bar{H}_0$ and one standard deviation width $\sigma_{H_0}$
by doing the integral
\begin{equation}
\label{eq:LikeHz} \mathcal{L}_H(\textbf{p}) = \frac{1}{\sqrt{2 \pi \sigma_{H_0}^2 }}\int_{\bar{H}_0-5\sigma_{H_0}}^{\bar{H}_0+5\sigma_{H_0}}e^{-\chi_{H}^2 (H_0, \textbf{p})/2}e^{-(H_0-\bar{H}_0)^2/(2 \sigma_{H_0}^2)}d H_0
\end{equation}
to get a likelihood function $\mathcal{L}_H(\textbf{p})$ that is a function of only the cosmological
parameters of interest (see, e.g., Ref. \cite{ Farooq_etal2013}). The best-fit  parameter values $p*$ are those
that maximize the likelihood function $\mathcal{L}_H(\textbf{p})$, or equivalently minimize $\chi_{H}^2 (\textbf{p})=-2 \mathcal{L}_H(\textbf{p})$.
The contours of 1, 2 and 3 $\sigma$ constraints  correspond to the sets of cosmological parameters (centered
on $p*$) bounded by $\chi^2_H(\textbf{p})=\chi^2_H(\textbf{p*})+2.3$, $\chi^2_H(\textbf{p})=\chi^2_H(\textbf{p*})+6.17$, and $\chi^2_H(\textbf{p})=\chi^2_H(\textbf{p*})+11.8$, respectively. It is worth to point out that the prior value of the Hubble constant $H_0$ significantly affects the cosmological parameter estimation~\cite{ Farooq_etal2013,Calabrese2012, Farooq_Ratra2013}.
In this work, we use the Gaussian prior on  $H_0$ directly from the Hubble Space Telescope (HST) observations of Cepheid variables
with $H_0 = 73.8 \pm 2.4$ km/s/Mpc~\cite{Riess2011}.

\begin{figure}[htbp]
\centering
\includegraphics[width=0.495\linewidth]{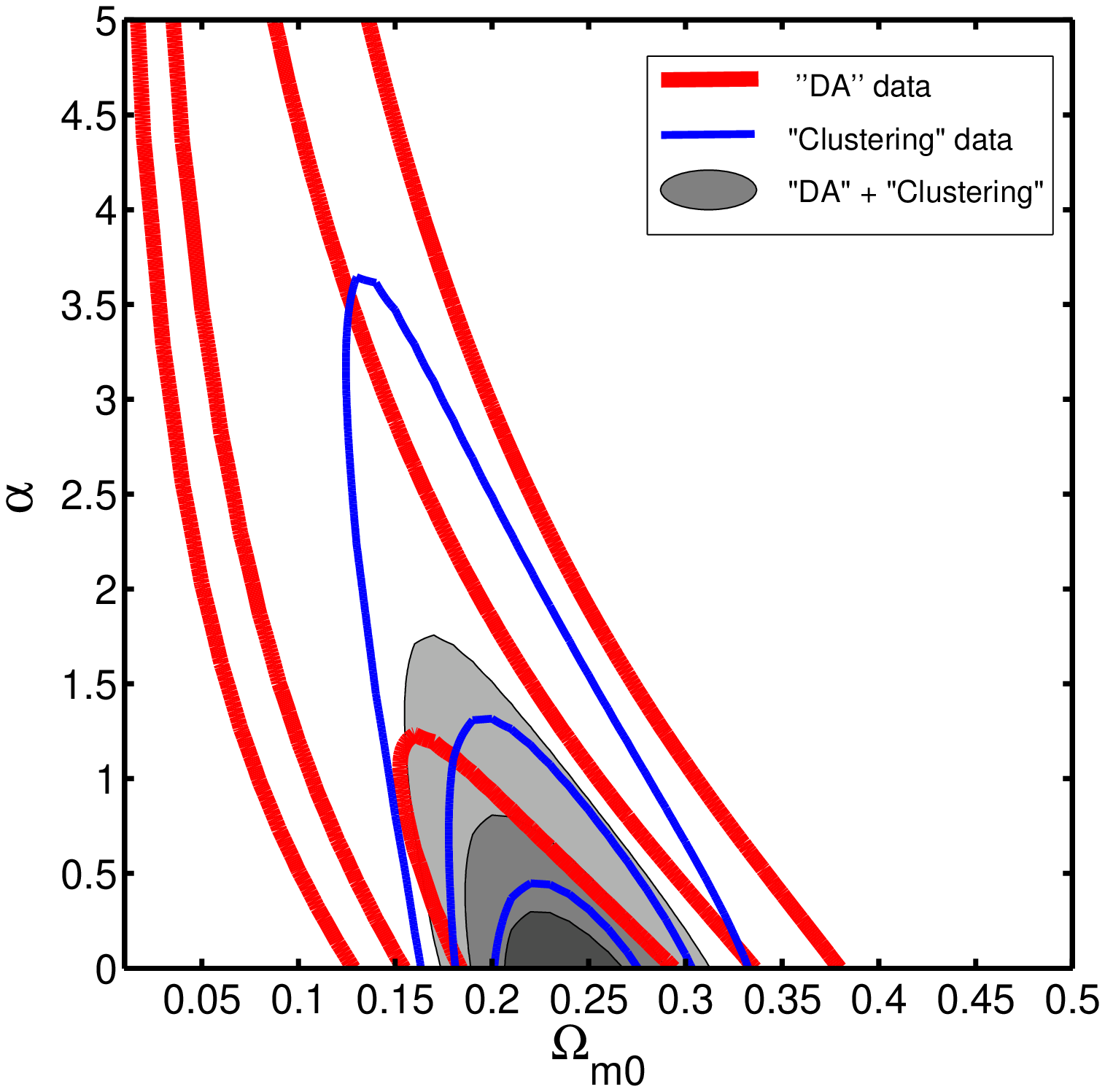}
\includegraphics[width=0.495\linewidth]{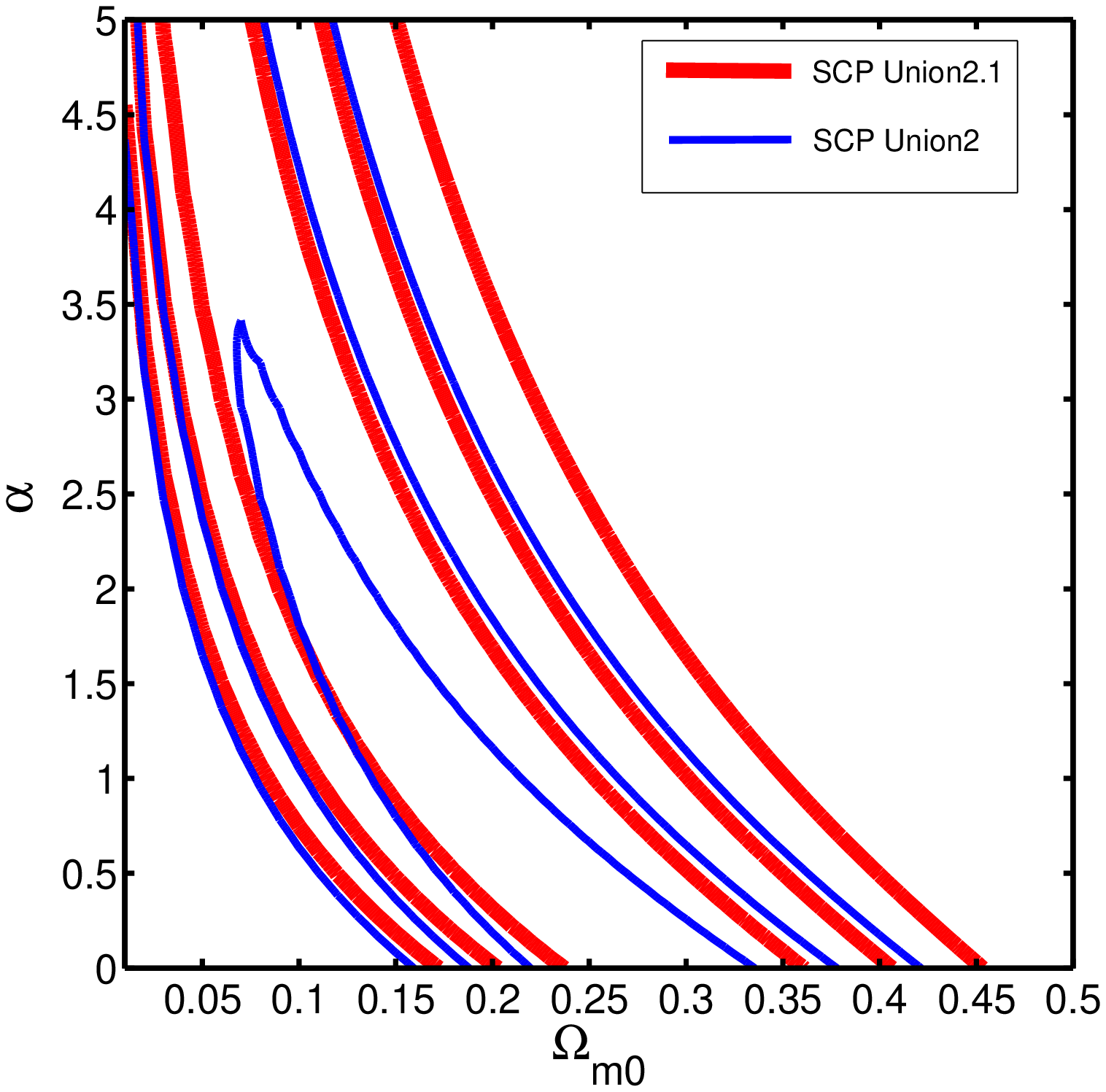}
\caption{Contours with 1, 2, and 3$\sigma$ constraints in the ($\Omega_{m0}$, $\alpha$) plane,
where the thick red lines, thin blue lines and gray shaded regions of the left panel correspond to
 constraints from the data of DA, Clustering and a combination of them, respectively,
 while the thin blue and thick red lines of the right panel
 correspond to the two recent SNe Ia samples, i.e., SCP Union2 and Union2.1  compilations, respectively. }
\label{fig:Hz_com}
   \end{figure}

Due to the differences between the ``DA'' and ``Clustering'' data,
we first study them separately as shown in the left panel of Fig.~\ref{fig:Hz_com}.
We see that the constraints from these two sub-samples
are consistent in 1$\sigma$,  implying that the systematical errors are subdominant in comparison to the statistical ones.
Furthermore, the constraints from ``Clustering'' data are tighter than those from the ``DA'' data, which can be understood
from Table~\ref{tab:Hz} as the uncertainties of  the ``DA'' data are generally larger than those of the ``Clustering''
ones. Then, by combining both the ``DA'' and ``Clustering'' data-sets, the results
are  presented as the gray shaded regions in the left panel of Fig.~\ref{fig:Hz_com}.
 Studies on the model from previous samples of the $H(z)$ data are also available from
 the literature~\cite{ Farooq_etal2013, Farooq_Ratra2013, Samushia2006, ChenY2011}.
 The results from the updated sample listed in the present work are significantly better than those from
 the much earlier samples with fewer data points~\cite{Samushia2006, ChenY2011}.
 We remark that, in Refs.~\cite{Farooq_etal2013, Farooq_Ratra2013} with the recent $H(z)$ data,
 two different prior values of $H_0$, namely, $H_0=68\pm 2.8 {\rm~km s^{-1} Mpc^{-1}}$ and
 $H_0=73.8\pm 2.4 {\rm~km s^{-1} Mpc^{-1}}$, are adopted. To eliminate the effect of the prior value of $H_0$,
 we can compare the constraints from our combined $H(z)$ sample with those shown in Fig. 3 of Ref.~\cite{Farooq_etal2013}
 and Fig. 3 of Ref.~\cite{Farooq_Ratra2013} for the same prior $H_0 = 73.8 \pm 2.4$ km/s/Mpc.
 We  note that the three correlated $H(z)$ measurements taken from Ref.~\cite{Blake2012} were simultaneously used
 in Ref.~\cite{Farooq_Ratra2013} without considering the covariance matrix. In addition, the $H(z=0.35)$ measurement from
 Ref.~\cite{Chuang2013} was mistakenly quoted as $H(z=0.35)=76.3 \pm 5.6$ km/s/Mpc.

In order to make a comparison, we display the constraints from the two recent SNe Ia samples, i.e., SCP Union2~\cite{Amanullah2010}
and Union2.1~\cite{Suzuki2012} compilations in the right panel of Fig.~\ref{fig:Hz_com}.
One finds that the constraints from the ``DA'' data are comparable to those from the SNe Ia samples.
However, those from the ``Clustering'' data and  the combined $H(z)$ sample are more restrictive
than the SNe Ia ones. In Refs.~\cite{Amanullah2010} and \cite{Suzuki2012}, SALT2~\cite{Guy2007} were used
to fit the supernova light curves. Note that in the SALT2 method there are three parameters: $x_0$, $x_1$, and $c$, corresponding to
the overall normalization to the time dependent spectral energy distribution (SED) of a SN Ia, the deviation  from the average
light curve shape, and the deviation from the mean SN Ia $\emph{B}-\emph{V}$ color, respectively.
We then combine
$x_1$, $c$ and the integrated B-band flux of  SALT2 SED at the maximum light $m^{max}_B$ to form the distance
modulus, given by
\begin{equation}
\label{eq:mu_obs} \mu^{\rm obs}_B(\alpha, \beta, M_B )=m_B^{max}-M_B+\alpha x_1-\beta c,
\end{equation}
where $M_B$ is the absolute B-band magnitude, and $\alpha$, $\beta$ and $M_B$ are nuisance parameters which are estimated simultaneously with the cosmological parameters and are marginalized over when obtaining the parameters of interest.
The theoretical (predicted) distance modulus is
\begin{equation}
\label{eq:mu_th} \mu^{\rm th}(z; \textbf{p}, \mu_0 )=5\log_{10} [D_L(z;
\textbf{p})]+\mu_0,
\end{equation}
where $\mu_0=42.38-5\log_{10}h$, which is also treated as a nuisance parameter, and the Hubble-free luminosity
distance is given by
\begin{equation}
\label{eq:DL} D_L(z;\textbf{p})=(1+z)\int_0^z
\frac{dz'}{E(z';\textbf{p})}.
\end{equation}
 The best-fit cosmological parameters from the SNe Ia data are determined by minimizing
\begin{equation}
\label{eq:chi2_SN1}
\chi^2_{SNe}=\sum_{i=1}^{N}\frac{\mu_B^{obs}(\alpha, \beta, M_B)-\mu^{th}(z_i;\textbf{p},\mu_0)}{\sigma_{ext}^2+\sigma_{sys}^2+\sigma_{lc}^2},
\end{equation}
where $N$ is the number of the data points, and a detailed discussion of  this equation can be found in
Refs.~\cite{Amanullah2010,Suzuki2012}. One can refer to Ref.~\cite{Giostri2012} for the methodology and process of marginalizing over the nuisance parameters.

\subsection{Recent strong gravitational lensing data}

In the framework of general relativity, the presence of matter can curve spacetime, so that the path of a light ray will be deflected.
This process is called gravitational lensing. The strong gravitational lensing (SGL) occurs when the source, lens and  observer are
so well aligned that the observer-source direction is close to or smaller than the so-called Einstein ring of the lens.
In the cosmological context, the source can be a quasar or  galaxy with a galaxy or  galaxy cluster acting as the lens. We will be concerned with both the lensing galaxy
and  cluster systems. In this paper we will use different methods to analyze the data associated to each of these two sets of systems. In the SGL systems, there are easily visible distortions such as the formation of multiple images, arcs or
even Einstein rings.

We first describe the analysis of the galaxy scale systems. It is known that in an axisymmetric lens, multiple images can only form in the vicinity of the so-called Einstein ring at an angle
$\theta_E$ from the center of the lens~\cite{Schneider1992}. Most of the lensing analyses have the Einstein radius ($\theta_E$)
as a basic quantity. In this work, we assume that the lens galaxies can be represented by the singular isothermal sphere (SIS)
or singular isothermal ellipsoid (SIE) models.
Moreover, the ellipticity in the lens galaxy mainly affects the relative numbers of two- and four-image lenses
but not their overall measurements of $\theta_E$.
The critical radius (Einstein radius) in the SIS lens or its SIE equivalent is given by
\begin{equation}
\label{eq:thetaE} \theta_E=4\pi \frac{\sigma_{SIS}^2}{c^2}\frac{D_{ls}}{D_s},
\end{equation}
which depends on the specific cosmological model,
where $D_{ls}$ is the
distance between the lens and source  and $D_s$
the observer and source. Note that in the SIS model, the separation between lensed images is always $2\theta_E$~\cite{Fukugita1992}.
The velocity dispersion $\sigma_{SIS}$ of the mass distribution and the observed stellar  velocity dispersion $\sigma_0$ do not have
to be the same.
By following Refs.~\cite{Kochanek1992, Ofek2003, CaoS2012}, we adopt a parameter $f_E$ to relate $\sigma_{SIS}$ and $\sigma_0$:
\begin{equation}
\label{eq:fE} \sigma_{SIS}=f_E \sigma_0,
\end{equation}
where $f_E$ is kept as a free parameter to mimic the effects of several systematic errors (see Refs.~\cite{Ofek2003, CaoS2012} for detailed discussions) from
(i) those in the root-mean-square (rms) difference between $\sigma_0$ and $\sigma_{SIS}$,
(ii) the rms error caused by assuming the SIS model in order to translate the observed image separation into $\theta_E$
as the observed image separation does not directly correspond to $\theta_E$,
and (iii) soften isothermal sphere potentials which tend to decrease the typical image separations~\cite{Narayan1996}.
The parameter $f_E$ is expected to be $(0.8)^{1/2}<f_E<(1.2)^{1/2}$~\cite{Ofek2003}.
In the method used in this paper, the cosmological model enters not through a distance measure directly, but rather  a distance ratio
\begin{equation}
\label{eq:Dth}
\mathcal{D}^{\rm th}(z_l,z_s;\textbf{p})=\frac{D_{ls}}{D_s}=\frac{\int_{z_l}^{z_s} \frac{dz}{E(z;\textbf{p})}}{\int_0^{z_s} \frac{dz}{E(z;\textbf{p})}},
\end{equation}
where in the framework of the flat FLRW metric the theoretical values of $D_s$ and $D_{ls}$ can be obtained by
\begin{equation}
\label{eq:Dsth} D_s(z_s; \textbf{p}, H_0) = \frac{c}{H_0(1+z_s)} \int_0^{z_s} \frac{dz}{E(z;\textbf{p})},
\end{equation}
and
\begin{equation}
\label{eq:Ddsth} D_{ls}(z_l, z_s; \textbf{p}, H_0) = \frac{c}{H_0(1+z_s)} \int_{z_l}^{z_s} \frac{dz}{E(z;\textbf{p})},
\end{equation}
respectively, where $c$ is the velocity of light and $z_{l(s)}$ is the redshift of the lens (source).
With Eq.~(\ref{eq:thetaE}), one can obtain the corresponding observable counterpart
\begin{equation}
\label{eq:Dobs}
\mathcal{D}^{\rm obs}_{LG}(f_E)=\frac{c^2 \theta_E}{4\pi \sigma_0^2 f_E^2},
\end{equation}
with its uncertainty calculated through the propagation equation concerning the errors on
$\sigma_0$ and $\theta_E$, where the Einstein radius $\theta_E$ is obtained from image astrometry,  the stellar velocity dispersion $\sigma_0$ from spectroscopy, and ``LG'' is short for ``Lensing Galaxy''. This method is independent of the Hubble constant (which gets canceled in the distance ratio) and  not affected by dust absorption or source evolutionary effects.
However, it depends on the measurements of $\sigma_0$ and $\theta_E$. In practice, the estimation of the Einstein radius $\theta_E$ from observed image positions depends on the lens modeling (e.g. the SIS or SIE assumption).
Moreover, the image separation could be affected by nearby masses (satellites, neighbor galaxies) and the structure along the line of sight, etc. These effects can lead to a $\thicksim$ 5\% systematic uncertainty in $\theta_E$~\cite{Grillo2008}. Additionally, from various sources of the lens system, we can roughly say that the error of the velocity dispersion is about 5\% ($\Delta \sigma_0/\sigma_0\thicksim 5\%$)~\cite{Lee2007}.
Briefly speaking, in Eq.~(\ref{eq:Dobs}) the introduction of $f_E$ accounts for systematic errors associated with both $\sigma_0$ and $\theta_E$. Cosmological model parameters constrained from the distance ratio $D_{ls}/D_s$ data of the lensing galaxy  systems can be estimated by minimizing the chi-square:
\begin{equation}
\label{eq:chi2GL} \chi^2_{LG}(\textbf{p}, f_E)=\sum_i \frac{(\mathcal{D}^{\rm obs}_{LG,i}(f_E)-\mathcal{D}^{\rm th}_i(z_l,z_s;\textbf{p}))^2}{\sigma_{\mathcal{D},i}^2}\,,
\end{equation}
which depends on the model parameter $\textbf{p}$ as well as  the nuisance parameter $f_E$.
We assume that the distribution of $f_E$ is a uniform function, $P(f_E)$, with the flat prior
$(0.8)^{1/2}<f_E<(1.2)^{1/2}$~\cite{Ofek2003}, and  marginalize over $f_E$ by doing the integral
\begin{equation}
\label{eq:LikelihoodGL} \mathcal{L}_{LG}(\textbf{p})=\int e^{-\chi^2_{LG}(\textbf{p}, f_E)/2} P(f_E)df_E.
\end{equation}
In practice, we maximize the likelihood $\mathcal{L}_{LG}(\textbf{p})$, or equivalently minimize $\chi^2_{LG}(\textbf{p})=-2\ln \mathcal{L}_{LG}(\textbf{p})$, with respect to the parameter  $\textbf{p}$ to find the best-fit parameters.
We consider a combined sample of 70 galaxy-lens systems with quasars or galaxies as sources 
from Sloan Lens ACS (SLACS) and Lens Structure and Dynamics (LSD) 
surveys~\cite{Grillo2008, Biesiada2010, Bolton2008, Grillo2009, Treu2003, Lehar1993, Tonry1998, Newton2011, WangandXu2013}. 
These 70 data points were listed in Table 1 of Ref.~\cite{CaoS2012}. It is worth to point out that the values of
$\mathcal{D}^{obs}$ listed in Ref.~\cite{CaoS2012} are obtained with $f_E=1$. 
In other words, they are in fact  the values of $\frac{c^2 \theta_E}{4\pi \sigma_0^2}$.  
We select the data points from the 70 ones with the selection criterion of $D_{ls}/D_s<1$, which means that
the distance between the lens and  source should be always smaller than that between the source and observer~\cite{LiaoK2012}. 
Note that the condition $D_{ls}/D_s<1$ should be satisfied in the whole range of values allowed for $f_E$. 
With the flat prior $0.8<f_E^2<1.2$, we choose the systems satisfying $\frac{c^2 \theta_E}{4\pi \sigma_0^2 f^2_E}<1$ when $f_E^2=1.2$.
In practice, 68 data points are picked out and  employed in our analyses.

We now describe the analysis of the cluster scale systems. Galaxy clusters, as the largest gravitationally bound systems  known in the universe, have their unique advantages for cosmology.
Strong lensing by galaxy clusters with galaxies acting as sources can produce giant arcs around the clusters,
which are perfect indicators of their surface mass densities, while the mass distributions of
the cluster mass halos can be modeled from the X-ray luminosity and temperature.
As a result, one can derive an observational
value of the distance ratio $D_{ls}/D_s$ to constrain cosmological parameters~\cite{Sereno2002, Sereno2004}. When a galaxy cluster is relaxed enough,
the pressure of its hot gas can balance its self-gravity. In this case, the hydrostatic isothermal spherically symmetric
$\beta$-model~\cite{Cavaliere1976} can be used to describe the intracluster medium (ICM) number density profile:
\begin{equation}
\label{eq:n_e}
n_e(r)=n_{e0}(1+r^2/r^2_c)^{-3\beta_X/2},
\end{equation}
 where $n_{e0}$ is the central electron number density, $r_c$ stands for the core radius, 
 and  $\beta_X$ describes the slope. The assumption of spherical symmetry is used everywhere else in the following content. 
 Under this assumption, one has
\begin{equation}
\label{eq:EulerEq}
-\frac{GM(r)}{r^2}=\frac{k_BT_X}{\mu m_p}\frac{d}{dr}\ln n_e(r),
\end{equation}
where $M(r)$ is the total mass of a cluster within radius $r$,
 $n_e(r)$ ($T_X$) is the gas  number density (temperature), 
 $k_B$ is the Boltzmann constant, $m_p$ is the proton mass, and $\mu$ denotes the mean molecular weight,
which is usually fixed with solar metallicity measurements, i.e., $\mu=0.585$ with an error of 4\%~\cite{Rosati2002}.
Inserting Eq.~(\ref{eq:n_e}) into Eq.~(\ref{eq:EulerEq}), one can get the cluster mass profile of the isothermal $\beta$-model,
\begin{equation}
\label{eq:Mr}
M(r)=\frac{3k_BT_X\beta_X}{G\mu m_p}\frac{r^3}{r^2_c+r^2}.
\end{equation}
The mass density distribution can be obtained with 
$\rho(r)=\frac{1}{4\pi}\frac{1}{r^2}\frac{dM(r)}{dr}$ \cite{Sereno2004}. 
Furthermore, the  mass density projected along the line of sight is 
\begin{equation}
\label{eq:SigmaTheta}
\Sigma(\theta)=\Sigma_0\frac{1+\frac{1}{2}(\frac{\theta}{\theta_c})^2}{[1+(\frac{\theta}{\theta_c})^2]^{3/2}},
\end{equation}
with $\Sigma_0$ the central surface density, given by
\begin{equation}
\label{eq:Sigma0}
\Sigma_0=\frac{3}{2}\frac{k_B}{G\mu m_p}\frac{T_X\beta_X}{\theta_c}\frac{1}{D_l},
\end{equation}
where $T_X$, $\beta_X$ and $\theta_c$ all come from the X-ray data fittings.
 In a spherically symmetric regular lens, a strong lensing event can occur if  \cite{Schneider1992}
 \begin{equation}
\label{eq:Sigma0_cr}
\Sigma_0>\Sigma_{cr},
\end{equation}
where $\Sigma_{cr}$ is the critical surface mass density,
\begin{equation}
\label{eq:Sigma_cr}
\Sigma_{cr}=\frac{c^2}{4\pi G}\frac{D_s}{D_l D_{ls}}.
\end{equation}
The locus in the image plane of  infinite magnification defines closed lines that do not intersect, which are called the ``critical lines'' \cite{Kneib2011}. A tangential critical curve appears at $\theta_t$ \cite{Schneider1992},
\begin{equation}
\label{eq:theta_t}
\theta_t=\theta_c\left[\left(\frac{\Sigma_0}{\Sigma_{cr}}\right)^2-1\right]^{1/2}.
\end{equation}
Due to the spherical symmetry, the tangential critical curve is just the Einstein ring, 
i.e., $\theta_t=\theta_E$. The value of $\theta_t$ can be derived from the observed arc radius $\theta_{arc}$.
However, according to Ref.~\cite{Ono1999}, the conventional analysis 
with $\theta_{t}=\theta_{arc}$ could overestimate the strong lensing mass by 10-30\%. To correct for this effect, we take $\theta_t=\epsilon \theta_{arc}$ \cite{Ono1999}, where the correction factor is $\epsilon=(1/\sqrt{1.2})\pm0.04$.
Observations of giant arcs in X-ray clusters enable us to estimate the distance ratio $D_{ls}/D_s$. 
Substituting Eqs.~(\ref{eq:Sigma0}) and  (\ref{eq:Sigma_cr}) into Eq.~(\ref{eq:theta_t}), one gets
\begin{equation}
\label{eq:DobsCL}
\mathcal{D}^{\rm obs}_{LC}=\frac{D_{ls}}{D_s}|_{obs}=\frac{\mu m_p c^2}{6\pi}\frac{1}{k_BT_X\beta_X}\sqrt{\theta_t^2+\theta_c^2},
\end{equation}
where ``LC'' is short for ``Lensing Cluster''.
The corresponding theoretical value is obtained from Eq.~(\ref{eq:Dth}). Recently, Yu \& Zhu~\cite{YuH2010}
collected a sample of 10 lensing galaxy clusters carefully selected from strong gravitational lensing systems which have
both X-ray satellite spectroscopic observations and optical giant luminous arcs with known redshifts. The sample satisfies the following
selection criteria. Firstly, the distance between the lens and  source should be always smaller than that
 the arc source and  observer, i.e., $D_{ls}/D_s<1$. Secondly, the arcs whose positions are too far
from the characteristic radius ($\theta_{arc}>3\theta_c$) should be discarded. Such cluster has a relatively small core radius and a much bigger arc radius. In this case, even X-ray observations cannot trace the matter at the region where lensing arcs are formed. As it is discussed in Sec. 2.3 of \cite{YuH2010}, the factor 3 is chosen empirically based on the observational data. The priors of all the necessary parameters included in
Eq.~(\ref{eq:DobsCL}) can be found in Table~1 of Ref.~\cite{YuH2010}. One can constrain the model with the sample of lensing clusters by minimizing the $\chi^2_{LC}$ function
\begin{equation}
\label{eq:chi2LC} \chi_{LC}^2 (\textbf{p}) =
\sum_{i=1}^{10}\frac{[\mathcal{D}^{\rm th}(\textbf{p})-\mathcal{D}^{\rm
obs}_{LC}]^2}{\sigma^2_{{\mathcal{D}},i}}.
\end{equation}

\begin{figure}[htbp]
\centering
\includegraphics[width=0.495\linewidth]{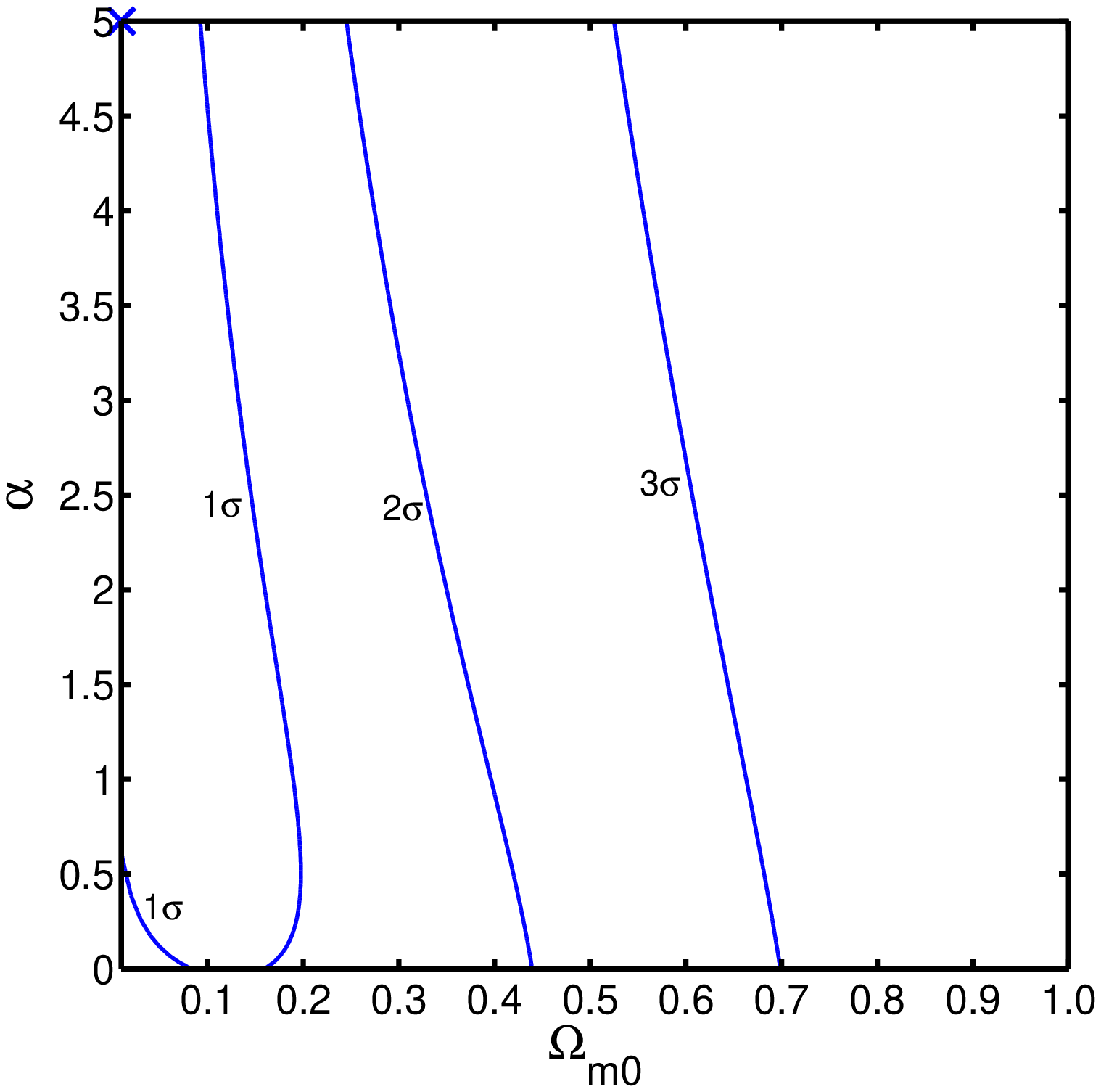}\\
\includegraphics[width=0.495\linewidth]{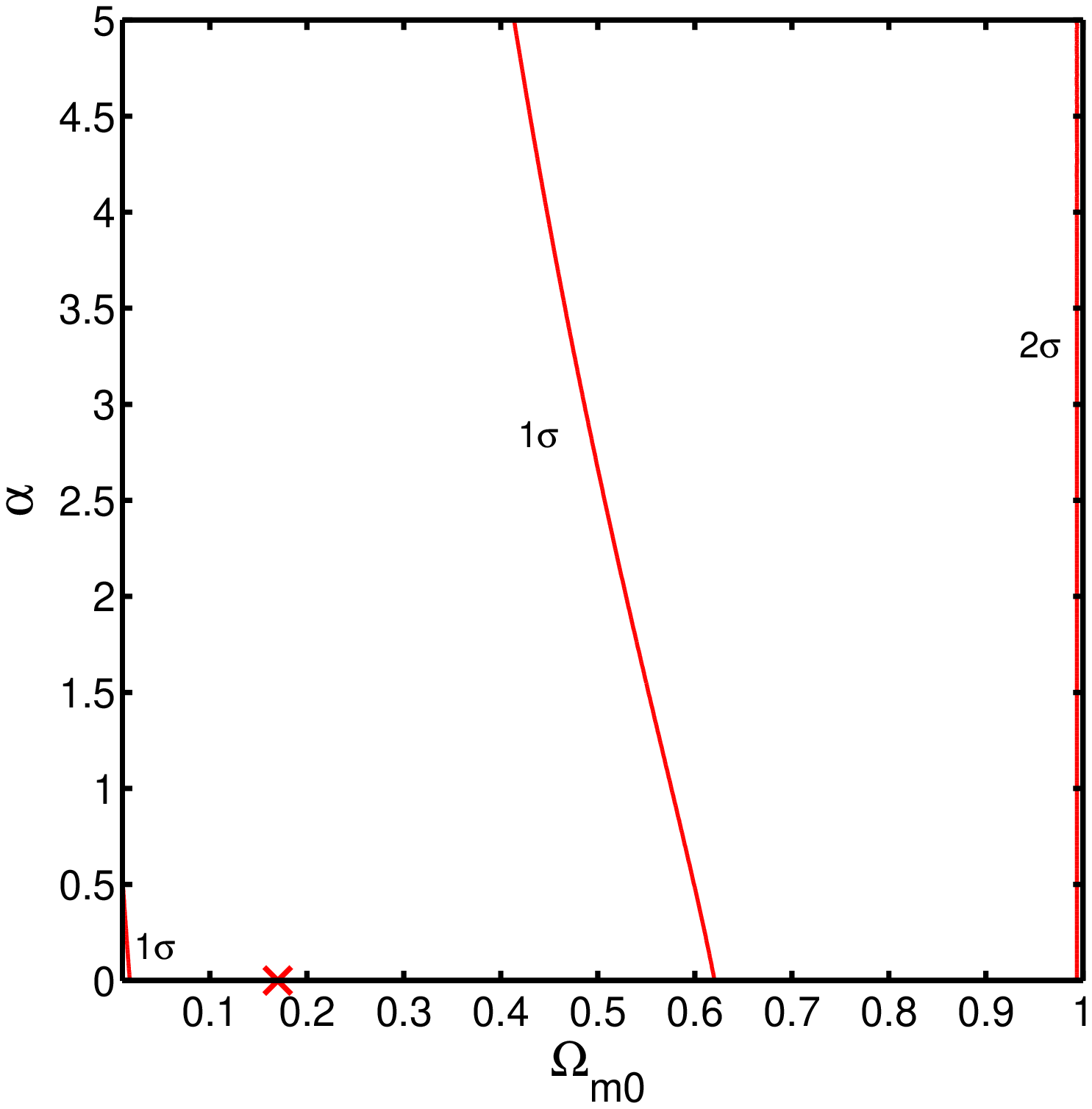}\\
\includegraphics[width=0.495\linewidth]{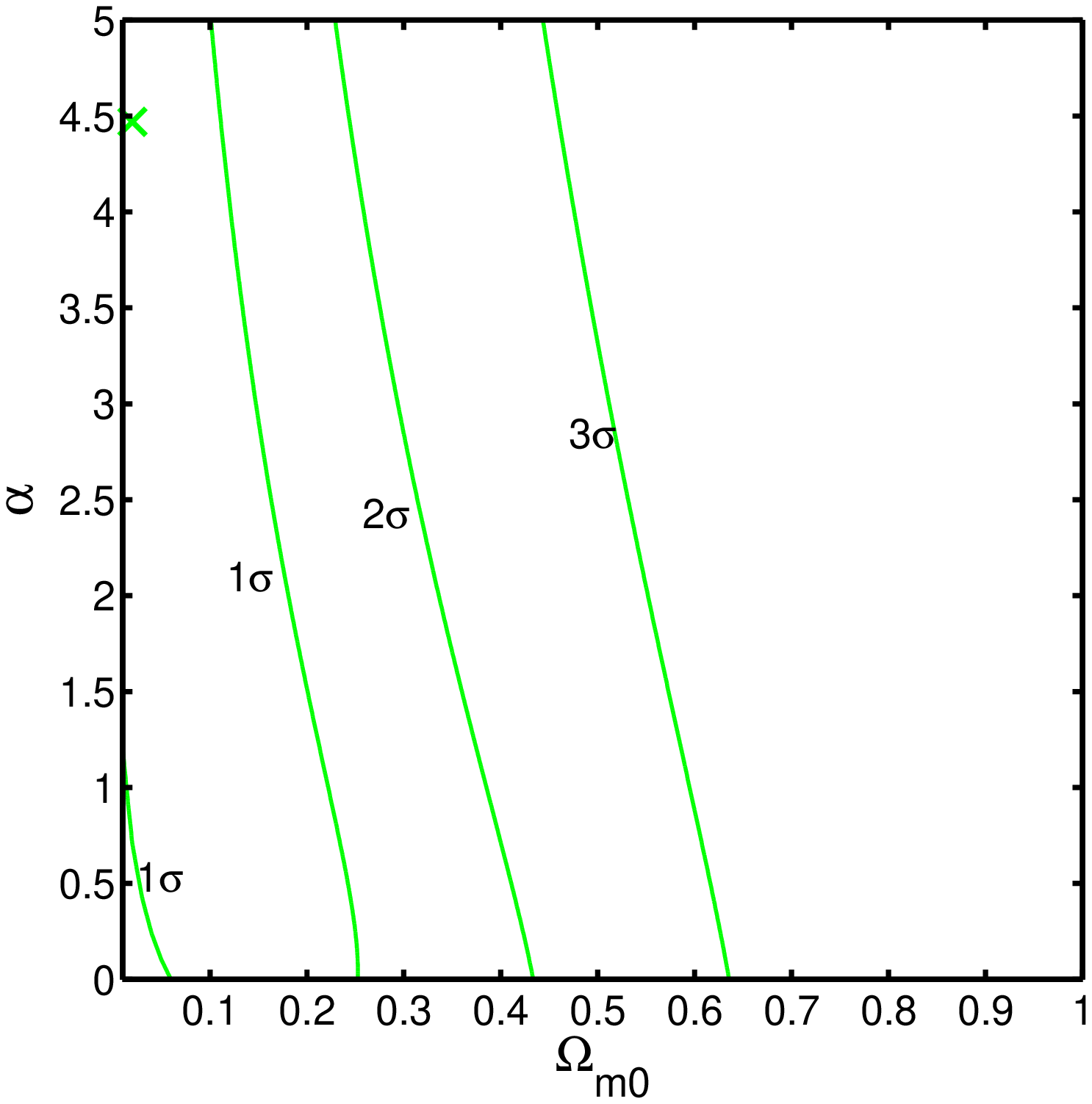}\\
\caption{Contours in the ($\Omega_{m0}$, $\alpha$) plane with
  1, 2 and 3 $\sigma$  representing $\Delta \chi^2$ values of 2.3, 6.17 and 11.8,
 where the top, middle and bottom panels correspond to the results from the 68 lensing galaxy,
 10 lensing cluster
 and the combination of lensing galaxy and lensing cluster systems,
 while  the three ``X'' symbols denote their best-fit values, respectively.
}
 \label{fig:SGL}
 \end{figure}

The top panel of Fig.~\ref{fig:SGL} illustrates constraints on the model parameters from the sample of 68 lensing galaxy systems.
The best-fit pair is $(\Omega_{m0}, \alpha)=(0.01, 5.0)$ with the minimum $\chi^2_{LG}=75.3$.
The results from the 10 lensing cluster systems are displayed in the middle panel of Fig.~\ref{fig:SGL}.
The best-fit pair is $(\Omega_{m0}, \alpha)=(0.17, 0.0)$ with the minimum $\chi^2_{LC}=13.5$.
The combination of the lensing galaxy and lensing cluster systems is
presented in the bottom panel of Fig.~\ref{fig:SGL}.  The relevant chi-square statistics is $\chi^2_{SGL}=\chi^2_{LG}+\chi^2_{LC}$, where the best fit is  $(\Omega_{m0}, \alpha)=(0.02, 4.47)$ with the minimum $\chi^2_{SGL}=89.2$.  It shows that the recent lensing galaxy and lensing cluster data sets cannot supply significant constraints on the cosmological parameters.

\subsection{Joint samples}

\begin{figure}[htbp]
\centering
\includegraphics[width=0.6\textwidth,height=0.6\textwidth]{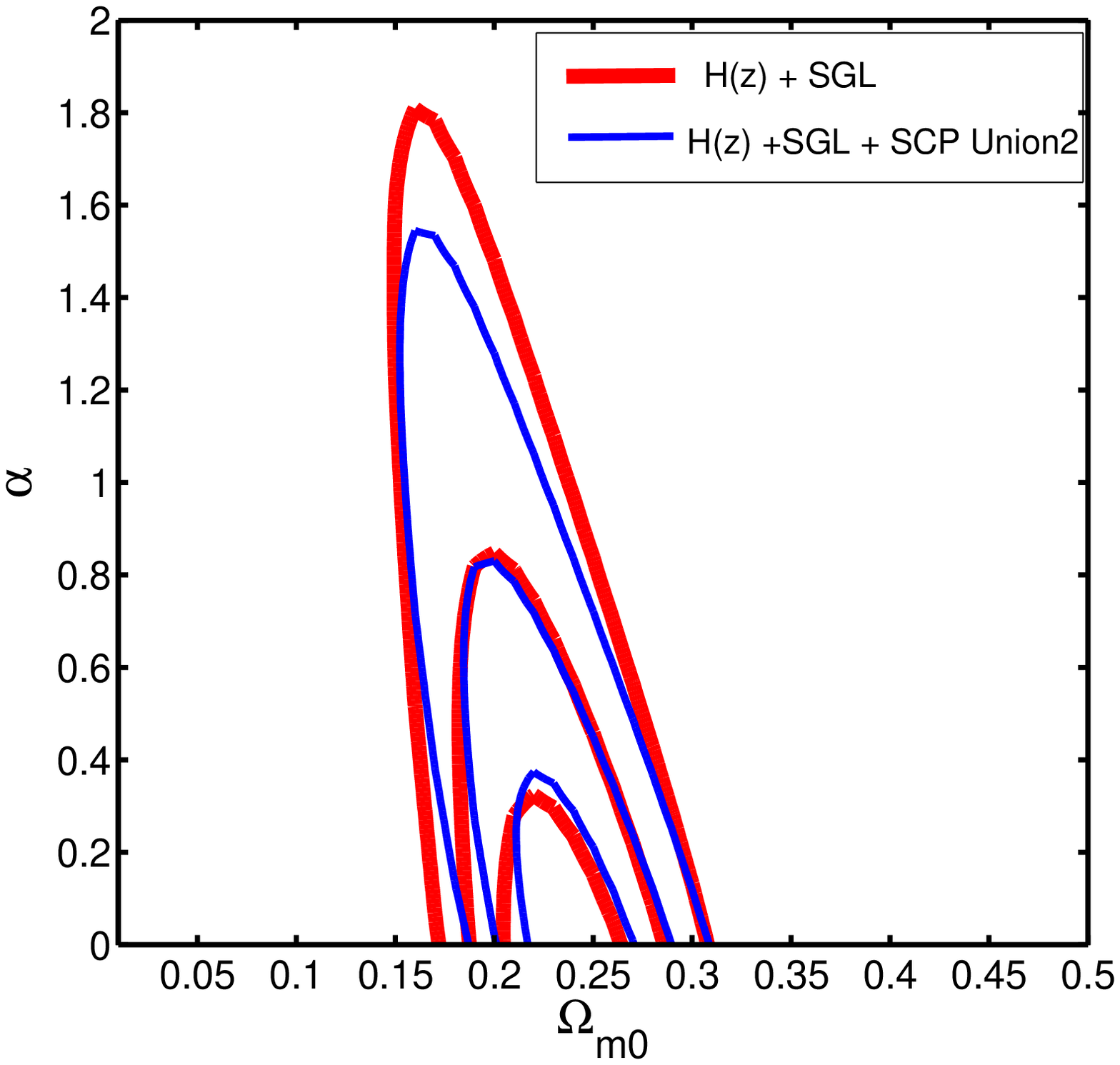}
\caption{
  Contours correspond to 1, 2 and 3 $\sigma$ in the ($\Omega_{m0}$, $\alpha$) plane from two joint samples.}
\label{fig:SGL_Hz_SN_com}
   \end{figure}

Fig.~\ref{fig:SGL_Hz_SN_com} illustrates our model parameters from different joint samples,
including a combination of the updated $H(z)$ (both ``DA'' and ``Clustering'') and SGL (both lensing galaxy and lensing cluster) data,
along with a combination of these two datasets and the SCP ``Union2'' compilation of SNe Ia. The intervals of the parameters constrained from different samples at 3 $\sigma$ confidence level are presented in Table~\ref{tab:JointConstraint}.
\begin{table*}
\caption{$\chi^2_{min}/d.o.f$ from different samples including bounds on the parameters $\Omega_{m0}$ and $\alpha$ at 3 $\sigma$ confidence level, where d.o.f denotes the degree of freedom.}\label{tab:JointConstraint}
\begin{center}
\begin{tabular}{lccl}
\hline
Sample & $\Omega_{m0}$ & $\alpha$  &$\chi^2_{min}/d.o.f$ \\
\hline
$H(z)$ data & $0.15<\Omega_{m0}<0.31$  & $0<\alpha<1.74$ &27.4/30\\
$H(z)$ data + SGL & $0.15<\Omega_{m0}<0.31$ & $0<\alpha<1.80$& 119.3/108 \\
$H(z)$ data + SCP Uinon2&  $0.16<\Omega_{m0}<0.31$ &$0<\alpha<1.47$ &559.1/587 \\
$H(z)$ data + SGL + SCP Uinon2& $0.15<\Omega_{m0}<0.31$ & $0<\alpha<1.54$&652.1/665  \\
\hline
\end{tabular}
\end{center}
\end{table*}

\section{Discussions and conclusions}
\label{summary}
We have concentrated on a quintessence model of dark energy with an inverse power-law potential $V(\phi)\propto \phi^{-\alpha}$ ($\alpha>0$),
in which the scalar field is taken as a candidate of dark energy to drive the late-time acceleration of the universe. This model is also
proven to alleviate the cosmological constant problems.
When $\alpha=0$, it is reduced to the corresponding $\Lambda$CDM model.

We have compiled and used a new sample of Hubble parameter measurements with 30 data points to constrain the parameters of the $\phi$CDM model. It turns out that the constraining power of the new $H(z)$ sample with the HST prior on $H_0$ is stronger than that of a recent SNe Ia sample. A recent SGL sample has also been employed to constrain the model parameters. However, the results indicate that constraints on the cosmological parameters from this sample are weak.  The observational data that we have studied here still can not distinguish whether dark energy is a time-independent cosmological constant or a time-varying dynamical component.
The complementary cosmological probes, such as the strong gravitational lensing and angular diameter distance measurements of galaxy clusters \cite{ChenY2012}, do not yet carry as much statistical weight as the SNe Ia, CMB, BAO and Hubble parameter data.

\section*{Acknowledgments}

We would like to thank Prof. Lixin Xu for helpful discussions. This
work was supported by the National Center for Theoretical Sciences, National
Science Council (Grant No. NSC-101-2112-M-007-006-MY3), National Tsing-Hua University (Grant No. 102N2725E1) at
Taiwan, the Ministry of Science and Technology
National Basic Science Program (Project 973) under Grant
No.2012CB821804, and the National Natural Science Foundation of China



\end{document}